# Non-contact Vital Signs Detection in Dynamic Environments

Shuai Sun
International College of Zhengzhou University
Zhengzhou, China
sun_23@stu.zzu.edu.cn

Chong-Xi Liang
Electronic Science and Engineering College of South China Normal University
Foshan, China
20223901059@m.scnu.edu.cn

Chengwei Ye
Homesite Group INC
Atlanta, USA
cye1@homesite.com

Huanzhen Zhang
Chewy
Atlanta, USA
hzhang2@chewy.com

Kangsheng Wang*
University of Science and Technology Beijing
Beijing, China
jackie@xs.ustb.edu.cn

*Abstract*—Accurate phase demodulation is essential for vital sign detection using millimeter wave radar. The time-varying DC offsets and phase imbalance in complex scenarios can seriously interfere with the performance of demodulation. This letter proposes a novel DC offset calibration algorithm as well as a Hilbert and differential cross-multiply (HADCM) demodulation algorithm to solve the time-varying imbalance terms. It works by estimating the time-varying DC offsets from neighboring peaks and valleys, and uses the differential form as well as the Hilbert transform of the *I/Q* channel signals to obtain the vital sign signal. Simulations and experiments have verified the effectiveness of the novel algorithm under low signal-to-noise ratio. Compared with the existing demodulation algorithms, the proposed algorithm can not only recover the original signal in complex environments more accurately, but also reduce the interference of noise on the signal.

*Keywords-component; DC offset calibration; FMCW radar; Hilbert and differential cross multiply demodulation algorithm; vital signs detection.*

## I. INTRODUCTION

In recent years, non-contact vital sign monitoring technology has been widely applied in fields such as smart homes [1], smart healthcare [2], driver monitoring systems [3], and post disaster search [4]. Among them, radar has received great attention in practical applications due to its advantages of low cost and high computational efficiency [5].

Accurate recovery of vital sign signals depends on the calibration and phase demodulation of *I/Q* signals [6]. DC offsets inevitably change due to variations in the environment. Arctangent demodulation (ATAN) method [7] is widely used to recover the phase. However, it is strongly influenced by DC offsets from the inevitable stationary clutter reflections. To recover the accurate phase information, various phase demodulation methods have been proposed based on the orthogonal characteristics of *I/Q* signals, including the extended differential cross-multiplication method (DACM) [8], modified DACM (MDACM) algorithm [9], enhanced arctangent (EATAN) algorithm [10], and amplitude-compensated complex signal demodulation (ACCSD) algorithm [11]. All of these algorithms assume that DC offsets remain constant during the detection of vital signs. The circle fitting algorithm has been used for removing DC offsets [12], [13]. However, for complex environments, time-varying DC offsets can lead to phase demodulation distortion [14]. In [15], a new adjacent chord angle accumulation (ACAA) algorithm is proposed, which makes the algorithm independent of DC offsets by accumulating the angles of adjacent chords of *I/Q* trajectory. However, due to the body's micromotion, the phase of *I/Q* channel signals will also inevitably vary. Therefore, a new method is needed that can solve the time-varying DC offsets problem while overcoming the variation in phase that affects the detection of *I/Q* channel signals.

In this paper, a novel phase demodulation algorithm is proposed. It estimates time-varying DC offsets by calculating peak-valley values of *I/Q* channel signals. In addition, the Hilbert transform and differential operation are introduced to perform Hilbert and differentiate cross-multiply (HADCM) algorithm on the calibrated signal, avoiding the impact of phase imbalance on demodulation caused by random body movements. Simulation and experimental results validate that this algorithm can work better under complex environment, compared with existing approach such as MDACM and ACAA algorithms.

## II. THEORY

Radar transmits electromagnetic waves through a transmitting antenna and gets the received signal after reflection from an object. The received signal is further analyzed to determine the location of the human target, and the vital sign signal is acquired by phase demodulation [16]. In the past research, only the constant phase error and DC offsets were considered. In order to obtain a more generalized algorithm, the time-varying amplitude, DC offsets, and the phase error are considered in the model. In a complex environment, the $I(t)$ and $Q(t)$ signals output from the mixer can be expressed as follows

$$I(t) = A_I(t)\cos(\theta + 4\pi\frac{d_0 + x(t) + \Delta d}{\lambda} + \varphi_{I_0}) + \mathrm{DC}_I(t), \quad (1)$$

$$Q(t) = A_Q(t)\sin(\theta + 4\pi\frac{d_0 + x(t) + \Delta d}{\lambda} + \varphi_{Q_0}) + DC_Q(t), \quad (2)$$

where $A_I(t)/A_Q(t)$ and $\varphi_{I_0}/\varphi_{Q_0}$ represents the amplitude and phase imbalance of I/Q channel signals, respectively. $\theta$ is the residual phase noise, which can be neglected. $d_0$ is the initial distance between the radar and the moving target. $x(t)$ represents the displacement of the chest wall. $\Delta d$ is the displacement due to the random body movement, normally less than 10 cm. To simplify the representation, $p(t) = 4\pi x(t)/\lambda$ denotes the vital sign signals to be detected. $\lambda$ is the wavelength of the received signal. $DC_I(t)$ and $DC_Q(t)$ are the time-varying DC offsets.

### A. Peak-Valley DC Offset Calibration Algorithm

Based on (1) and (2), the expression of I/Q channel signals when it obtains an extreme value in a short time is:

$$I_{\max}(t) = A_I(t) + DC_I(t), \quad I_{\min}(t) = -A_I(t) + DC_I(t), \quad (3)$$

$$Q_{\max}(t) = A_Q(t) + DC_Q(t), \quad Q_{\min}(t) = -A_Q(t) + DC_Q(t). \quad (4)$$

In this letter, an efficient peak-valley DC offset calibration algorithm is used for dynamic DC offset estimation. Fig. 1 shows the diagram of the proposed algorithm using $DC_I$ as an example. For I/Q channel signals, adjacent peaks and valleys can be considered to be satisfied with $t_{\max_{j(k)}} \approx t_{\min_{j(k)}}, j=1,2,...,m; k=1,2,...,n$, $m/n$ represents the number of peaks (valleys) achieved by the I/Q signal, respectively. Consequently, the time-varying discrete form's DC offsets can be written as follows

$$DC_I(j) = \frac{I_{\max}(t_{\max_j}) + I_{\min}(t_{\min_j})}{2}, \quad (5)$$

$$DC_Q(k) = \frac{Q_{\max}(t_{\max_k}) + Q_{\min}(t_{\min_k})}{2}. \quad (6)$$

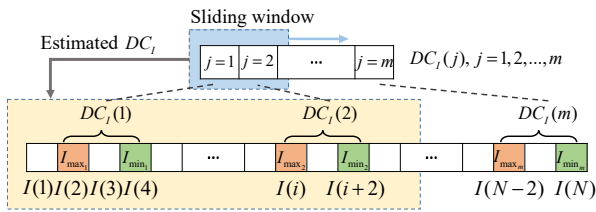

Figure 1. Simplified diagram of the peak-valley calibration algorithm.

Subsequently, a sliding window is chosen to calculate the average value of $DC_I(j)/DC_Q(k)$ over a short period of time, which is used as an estimation of $DC_I/DC_Q$ over this period. The calibrated I/Q channel signals can be expressed as

$$I_C(t) = A_I(t)\cos(p(t) + \varphi_I), \quad (7)$$

$$Q_C(t) = A_Q(t)\sin(p(t) + \varphi_Q), \quad (8)$$

where $\varphi_I$ and $\varphi_Q$ are the phase imbalance caused by I/Q channel imbalance and random body movements.

### B. Hilbert and Differentiate Cross-Multiply Algorithm

Implementing the Hilbert transformation and differential operation after removing DC offsets, which can be expressed as:

$$I_H(t) = A_I(t)\sin(p(t) + \varphi_I), \quad (9)$$

$$Q_H(t) = -A_Q(t)\cos(p(t) + \varphi_Q), \quad (10)$$

$$\hat{I}_C(t) = -A_I(t)\sin(p(t) + \varphi_I) \cdot \hat{p}(t), \quad (11)$$

$$\hat{Q}_C(t) = A_Q(t)\cos(p(t) + \varphi_Q) \cdot \hat{p}(t). \quad (12)$$

(9)-(12) can be cross-multiplied and combined with the trigonometric operations to get $I_C(t)Q_H(t)$, $I_H(t)Q_C(t)$, $I_C(t)\hat{Q}_C(t)$ and $\hat{I}_C(t)Q_C(t)$. Then, further operations can be performed as

$$I_H(t)Q_C(t) - I_C(t)Q_H(t) = A_I(t)A_Q(t)\cos(\varphi_I - \varphi_Q), \quad (13)$$

$$I_C(t)\hat{Q}_C(t) - Q_C(t)\hat{I}_C(t) = A_I(t)A_Q(t)\cos(\varphi_I - \varphi_Q)\hat{p}(t). \quad (14)$$

According to (13)(14), it is obvious that the derivative of the phase can be written as

$$\hat{p}(t) = \frac{I_C(t)\hat{Q}_C(t) - Q_C(t)\hat{I}_C(t)}{I_H(t)Q_C(t) - I_C(t)Q_H(t)}. \quad (15)$$

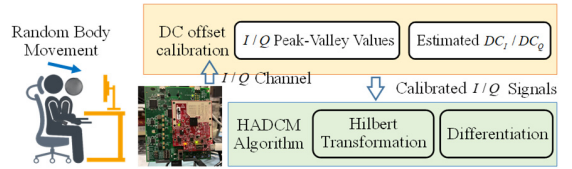

Figure 2. The flowchart of the proposed algorithm.

The flowchart of the proposed demodulation algorithm is shown in Fig. 2. After obtaining I/Q channel signals from the quadrature demodulated baseband signal, the peak-valley DC offset calibration algorithm is first used to get calibrated I/Q signals. Then, the real phase signal is recovered with the HADCM demodulation algorithm. It is worth noting that variations in $\varphi_I$ and $\varphi_Q$ due to random body movements have no effect on the detection of $p(t)$.

### III. SIMULATION

According to the vital sign model, the target movement is set as sinusoidal motion with added noise. The sampling rate is set as 20Hz. The respiratory frequency and cardiac frequency are set to 0.3 Hz and 1.3 Hz, respectively. The respiratory and cardiac amplitude are set to 6 mm and 0.3 mm, respectively. The imbalance term of the model includes phase errors of I/Q channel signals, which are usually set as $\varphi_I = \pi/12$ ,

$\varphi_Q = \pi/15$, time-varying DC offsets $DC_I$=random(1,3), $DC_Q$=random(1,3).

Here, $random(a,b)$ means a random number is selected from $(a,b)$. Additionally, the size of the sliding window also affects the DC offset estimation. Apart from the previously mentioned parameter settings, we also focused on the case of window lengths of 1s, 2s, 3s and 4s, with no overlapping, respectively.

To quantitatively characterize the effectiveness of different algorithms in removing DC offsets, the DC offset relative error $e_I / e_Q$ is defined as follows:

$$e_I = \frac{|I(t) - I_c(t)|}{DC_I}, \quad (16)$$

$$e_Q = \frac{|Q(t) - Q_c(t)|}{DC_Q}, \quad (17)$$

where $I(t)/Q(t)$ includes DC offset, $I_c(t)/Q_c(t)$ is the calibrated I/Q signal, and $DC_I / DC_Q$ represents the set DC offset.

TABLE I. SIMULATION RESULTS FOR DIFFERENT WINDOW LENGTH WITH CIRCLE FITTING METHOD AND THE PEAK-VALLEY CALIBRATION ALGORITHM

| Relative error | Method | Window length(s) | | | |
|---|---|---|---|---|---|
| | | 1 | 2 | 3 | 4 |
| $e_I(e_Q)$ | Circle fitting | 0.29(0.24) | 0.46(0.48) | 0.42(0.35) | 0.33(0.13) |
| | This work | 0.69(0.66) | 0.94(0.87) | 0.79(0.73) | 0.68(0.60) |

Table I shows simulation results for different window length conditions and the $e_I / e_Q$ calculated with the circle fitting method [12] and the proposed algorithm. Obviously, a larger $e_I / e_Q$ indicates that the estimated DC offset is closer to the true offset. On the contrary, there will be a significant deviation from the actual signal. It can be observed that peak-valley DC offset calibration algorithm is more effective in removing time-varying DC offsets and restoring the original signal compared with the circle fitting algorithm. In addition, the best results are obtained when the sliding window length is 2s.

In order to evaluate the demodulation performance of the HADCM algorithm in complex environments, simulations are carried out under different signal-to-noise ratios (SNRs). The DC offsets are first removed with the proposed algorithm. Subsequently, phase related to vital sign is extracted and compared with different algorithms. Fig. 3(a) and Fig. 3(b) show the results of demodulation using ATAN [7], MDACM [9], ACAA [15], and HADCM algorithm under different SNRs. In Fig. 3(a), for a SNR of 30 dB, ATAN demodulation produces a larger deviation, the signals obtained by MDACM and ACAA demodulation algorithms still have certain error with the ideal signal, but the HADCM demodulation results are basically consistent with the ideal signal. At a SNR of 10 dB, the ATAN, MDACM, and ACAA demodulation algorithms produce large deviation due to noise and the phase error, as shown in Fig. 3(b). In contrast, the HADCM algorithm is more effective.

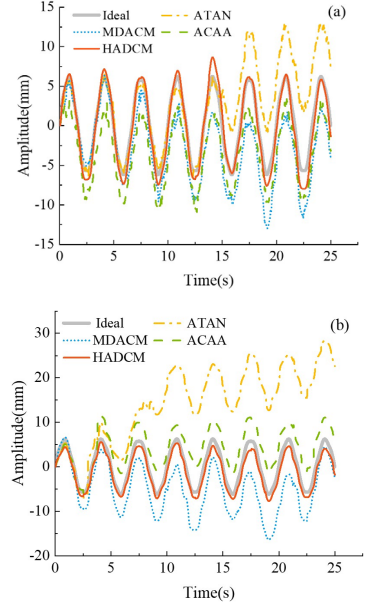

Figure 3. The demodulation results of p(t) with four algorithms under SNR=30dB and 10dB. (a) SNR=30db, (b) SNR=10dB.

TABLE II. RMSEs WITH DIFFERENT DEMODULATION ALGORITHMS UNDER DIFFERENT SNRs

| SNR (dB) | ATAN(mm) | MDACM(mm) | ACAA(mm) | HDACM(mm) |
|---|---|---|---|---|
| 10 | 16.02 | 5.93 | 4.54 | 1.33 |
| 15 | 14.24 | 5.32 | 4.34 | 1.21 |
| 20 | 10.22 | 4.25 | 4.12 | 1.03 |
| 25 | 8.54 | 3.93 | 3.82 | 0.99 |
| 30 | 4.12 | 3.74 | 3.66 | 0.96 |

To compare the accuracy of the reconstructed signals with different algorithms more accurately, the root mean square error (RMSE) is used to describe the error. Table II shows the RMSEs after demodulation using the above algorithms as the SNR changes from 10 dB to 30 dB in step of 5 dB. It is clear to see that the RMSE of the signal demodulated by the ATAN algorithm decreases rapidly as the SNR increases. Meanwhile, the RMSE of the signal demodulated by MDACM and ACAA varies very little, but remains stable above 3.5 mm. In contrast, the HADCM algorithm has stronger noise robustness compared to other algorithms.

## IV. EXPERIMENTS

Texas Instruments (TI) 60GHz IWR6843 millimeter wave radar was used in the experiment. The radar's sampling rate was set to 20Hz. The bandwidth was 600MHz. In the first experiment, the tester was located in a complex office environment. Unlike the first experiment, the second experiment included another person who walked back and forth at various distances nearby, which severely interfered with the detection of the tester's vital signs. The test subject was located 60cm away from the radar, kept still for 40 seconds and swayed his body back and forth slightly for 20 seconds, without exceeding the range resolution. Moreover, the respiratory rate was controlled at 14 breaths per minute, with the consent of the tester.

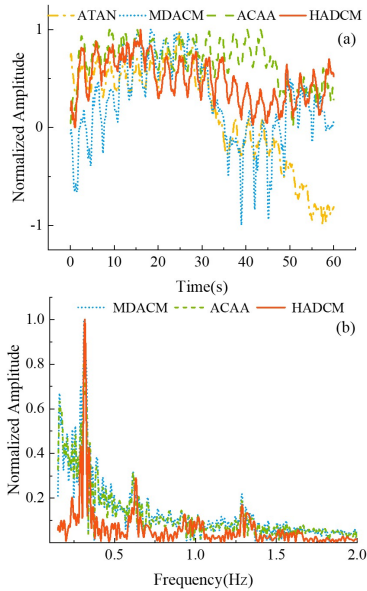

Figure 4. Comparison of the experimental results with different algorithms: (a) Time domain (b) Frequency domain.

As for the first experiment, Fig. 4(a) shows the vital sign signal after demodulation. It is clear to see that the demodulated signals with the ATAN and MDACM algorithms produce severe distortion. Although the ACAA demodulation algorithm can avoid time-varying DC offsets, it is unable to recover vital sign signals when there is random body movement (40s-60s). However, these problems can be overcome with the HADCM algorithm. Fig. 4(b) shows the normalized spectrum of the demodulated signal using different algorithms. It is obvious that the signal demodulated by the HADCM algorithm contains less noise. The mean errors of respiration rate (RR) detection for experiment 2 are shown in Table III. As the distance between the interfering person and the tester decreases, the interference with the detection of the tester's vital signs increases. It can be seen that the HADCM algorithm achieves the best results, with the mean RR error controlled below 2 beats per minute (bpm). Therefore, the effectiveness of the HADCM algorithm in detecting vital signs in complex environments is verified.

TABLE III. THE MEAN RR ERRORS FOR INTERFERENCE DIFFERENT DISTANCES

| Distance (cm) | ATAN (bpm) | MDACM (bpm) | ACAA (bpm) | HADCM (bpm) |
|---|---|---|---|---|
| 100 | 5.82 | 3.11 | 0.56 | 0.17 |
| 80 | 6.31 | 5.48 | 3.16 | 1.14 |
| 60 | 8.53 | 7.22 | 4.07 | 1.87 |

## V. CONCLUSIONS

This paper presents a peak-valley DC offset calibration algorithm and HADCM algorithm for vital sign detection in complex environments. It can better address issues including time-varying DC offsets and phase error in demodulation. Furthermore, the feasibility of this algorithm is verified by simulations and experiments. This work provides a solution for the detection of non-contact vital sign in the presence of random body movement.